\documentclass[a4paper,twocolumn, nofootinbib,superscriptaddress,aps,prl, 14pt,eqsecnum,notitlepage,showkeys]{revtex4-1}
\usepackage{multirow}
\usepackage{graphicx}
\usepackage{bm}
\usepackage{dcolumn}
\usepackage{hyperref}
\usepackage{gensymb}
\usepackage{amsmath}
\usepackage{dcolumn}    
\usepackage{ifpdf}
\usepackage{amssymb,lineno,amsfonts}
\usepackage{graphicx}   
\usepackage{bm}         
\usepackage{bbm}
\usepackage{mathrsfs}
\usepackage{upgreek}
\usepackage{mathtools}
\usepackage{epstopdf}
\usepackage{setspace}
\usepackage{hyperref}
\usepackage{natbib}
\usepackage{esvect}
\usepackage{amsmath}
\usepackage[usenames,dvipsnames]{xcolor}
\definecolor{med-blue}{RGB}{25,25,112}
\hypersetup{colorlinks, linkcolor={blue},citecolor={blue}, urlcolor={blue}}
\usepackage{times }
\usepackage [english]{babel}
\usepackage [autostyle, english = american]{csquotes}
\MakeOuterQuote{"}
\begin{document}	
	
	\title{Dielectric and Raman Spectroscopy in Hematite crystallites across the Morin Transition}
	
	\author{Namrata Pattanayak}
	\author{Jitender Kumar}
	\author{Partha Pratim Patra}
	\affiliation{Department of Physics, Indian Institute of Science Education and Research, Dr. Homi Bhabha Road, Pune 411008, India}
	\author{G. V. Pavan Kumar}
	\author{Ashna Bajpai}
	\affiliation{Department of Physics, Indian Institute of Science Education and Research, Dr. Homi Bhabha Road, Pune 411008, India}
	\affiliation{Center for Energy Science, Indian Institute of Science Education and Research, Dr. Homi Bhabha Road, Pune 411008, India}

	\begin{abstract}
		We report complex dielectric and Raman spectroscopy measurements in four samples of $\alpha$- Fe$_2$O$_3$, consisting of crystallites which are  either hexagonal shaped plates or cuboids. All four samples exhibit the spin reorientation transition from a pure antiferromagnetic (AFM) to a weak-ferromagnetic (WFM)  state at the Morin Transition temperature (T$_M$) intrinsic to $\alpha$- Fe$_2$O$_3$. These samples, pressed and sintered in identical conditions for the dielectric  measurements, reveal moderate but clear enhancement in the real part of the dielectric constant ($\epsilon'$) in the WFM region. However, a relaxation-like behavior in the imaginary part of  $\epsilon''$ is observed  only in nano plates or big cuboids.  Further still, this relaxation patten is observed only in lower frequency region, lasting upto a few kHz and follows Arrhenius law within this limited range. The activation energy deduced from the fitting is suggestive of polaronic conduction. Temperature dependent Raman spectra reveal anomalies in all major phononic modes and also in 2Eu mode in the vicinity of the Morin transition.  A peak like behavior in Raman  Shifts, in conjuncture  with  anharmonic fitting reveals that  the nature of spin phonon coupling is different in pure AFM and WFM regions and it is tied to the mild variations, as observed in the dielectric constant of$\alpha$- Fe$_2$O$_3$ near the T$_M$.    
	\end{abstract}
	
	%
	%
	%
	\maketitle
	%
	%

\section{Introduction}

Materials exhibiting an entanglement between magnetic and electric order parameters are well known for their tremendous technological importance in futuristic memory storage and spintronic devices \cite{Ortega,Ramesh,Nicola}. For these systems, stringent symmetry limitations are required to realize the coupling between electric polarization (P) and magnetization (M).  Cr$_2$O$_3$, for instance  is a prototypical magnetoelectric (ME) compound that exhibits linear coupling between P and M. \cite{Dzy1,Astrov}. For magnetic systems with non collinear spin structure such as TbMnO$_3$, the appearance of nonzero polarization is understood to arise from  the antisymmetric Dzyaloshinskii-Moriya Interaction (DMI) and such a coupling is again governed by very specific symmetry requirements related to the magnetic lattice. \cite{Aoyama,Kimura1,Hur,Katsura}. Magnetodielectric (MD) materials, on the other hand,  exhibit anomalies in dielectric constants across the magnetic transition temperatures, either ferromagnetic (FM) or antiferromagnetic (AFM) ones.  This is irrespective of a specific magnetic symmetry  or the peculiar spin configuration such as involved in ME or other DMI driven canted systems.\cite{Lawes,Tackett,Kimura2,Newnham}. The dielectric  anomalies in MD  materials are tied to spin-phonon interactions and such systems (single phase or composites)  are therefore equally important for practical applications \cite{Shin,Tiwari,Singh}.

	\begin{table*}[h]
  \centering
  \begin{tabular}{|c|c|c|c|c|c|c|}
  \hline \hline
\textbf{Sample} & \textbf{S/V} & \textbf{Morin} &\multicolumn{3}{c|}{\textbf{Lattice constant}} \\
 & \textbf{ratio} & \textbf{Temperature} &\multicolumn{3}{c|}{} \\\cline{4-6}   
              & $nm^{-1}$ & (K) & c (\AA) & a (\AA) & c/a \\\cline{1-6}
Micro Plates  & 0.008 & 225 & 13.7586(4)  & 5.0386(1) &  2.7306(1)\\\cline{1-6} 
Nano Plates   & 0.16 & 172 & 13.7593(3)   & 5.0375(0) & 2.7313(9)\\\cline{1-6} 
Big Cuboids   & 0.03 & 210 & 13.7678(7)& 5.0371(2) & 2.7332(1)  \\\cline{1-6}
Small Cuboids & 0.1 & 185 & 13.7687(1) & 5.0350(0) &  2.7346(2)  \\\cline{1-6}
    
    \hline \hline   
    \end{tabular}
     \caption{\label{Table}The S/V ratio, Morin transition temperature and structural parameters of $\alpha$- Fe$_2$O$_3$ samples as determined from the Rietveld analysis of the room-temperature x-ray diffraction data.}
\end{table*}
	
On a general note, it is non trivial to establish the exact origins of dielectric anomalies and their coupling with the magnetic order either in symmetry allowed ME systems, in routine FM and AFM or in canted AFM. This is due to the presence of a number of factors, both intrinsic and extrinsic which influence the dielectric constant at the magnetic transition temperature \cite{Lawes,Lunkenheimer, Catalan} etc. These studies also provide the pointers, through frequency and magnetic field dependent dielectric spectroscopy, to establish or isolate various factors which may  contribute to the  dielectric anomalies especially around the magnetic transition. This includes  the role of surface defects, oxygen vacancies and  magnetoresitive contributions etc.\cite{Lawes}.

 In the context of DMI driven coupling between magnetic and electrical order parameters, \cite{Khomskii}  $\alpha$- Fe$_2$O$_3$   is unique, as it exhibits  both pure  AFM as well as spin  canted phase. More importantly the canted phase exists at the room temperature.\cite{Dzy2,Moriya1,Khomskii}.Below its Neel temperature (T$_N$ $\sim$ 950 K) the AFM sublattices of $\alpha$- Fe$_2$O$_3$  lie in the basal plane of hexagonal structure (a convenient representation $\alpha$- Fe$_2$O$_3$) and slightly canted, resulting in a WFM phase. With the decrease of temperature, a  spin reorientation transition occurs, known as Morin transition (T$_M$ $\sim$260 K). AT T$_M$,  the sublattice magnetization reorients from the ab plane to the c-axis of the unit cell.  This phenomenon is accompanied with  simultaneous vanishing of the DMI driven spin canting and  $\alpha$- Fe$_2$O$_3$ becomes a pure AFM. In addition to being a room temperature  WFM, $\alpha$- Fe$_2$O$_3$ is also a symmetry allowed piezomagnet,  a phenomena that involves the generation of magnetic moments  upon application of stress \cite{Dzy3, Andratskii}.

\begin{figure*}[!t]
\includegraphics[width=1\textwidth]{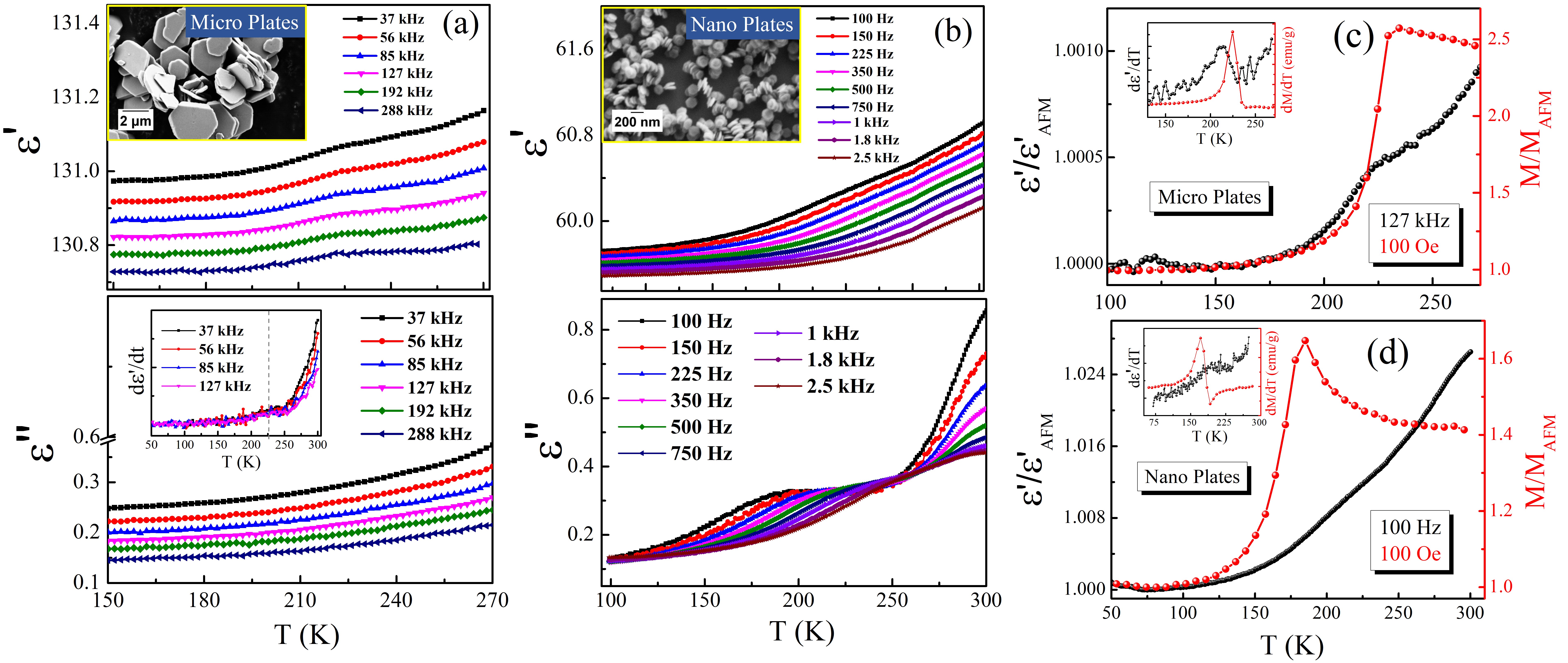}
\caption {Temperature variation of $\epsilon'$ (upper panel) and $\epsilon''$ (lower panel) in (a) Micro plates, (b) Nano plates. (c) an (d) are normalized  $\epsilon'$ in conjunction with the normalize Magnetization data for the micro and  and nano plates respectively. The insets in (c) and (d) depicts the temperature derivatives of $\epsilon'$ and magnetization indicating dielectric anomaly occurring in the vicinity of T$_M$ in both the samples.}
\label{Figure1}
\end{figure*}

     DMI driven anomalies in dielectric constant are also subject matter of investigation in various magnetic insulators \cite{Khomskii}.  However, to the best our knowledge, how this feature  influences the dielectric constant in prototypical DMI driven compound $\alpha$- Fe$_2$O$_3$, which is also room temperature WFM, has not been explored in great details.  Although the ME coupling  is not expected in $\alpha$- Fe$_2$O$_3$, a close inspection of the DMI driven coupling on the dielectric constant, especially its variations with morphology and nano structuring is interesting. In a recent study, a signature of magneto-dielectric coupling has been observed at the vicinity of Morin transition, T$_M$, in $\alpha$- Fe$_2$O$_3$. The observed anomaly is also seen to be enhanced with Ga doping in the system \cite{Lone}. However, in all these studies the temperature region of the maxima of dielectric anomaly and the Morin transition temperature are largely separated ( by $\sim$75 K). More importantly,  the anomaly is seen to be pronounced in doped $\alpha$- Fe$_2$O$_3$ system, albeit  less attention has been paid in understanding the nature of coupling phenomena in the pristine $\alpha$- Fe$_2$O$_3$, especially as a function of nano-structuring and morphology.  
		
In the present work we investigate dielectric characteristic  in four different samples of $\alpha$- Fe$_2$O$_3$ in its pure AFM and WFM regions. The samples are primarily formed in two different morphologies (i) hexagonal plates and (ii) cuboids. To separate the size effects, while retaining the morohology, hexagonal plates under investigation are referred as \textit{micro-plates}  and \textit{nano-plates} respectively. On the similar lines, cuboid have been reffed to as \textit{big cuboids}  and \textit{small cuboids} in this work. In addition to dielectric spectroscopy, we have also performed Raman spectroscopy as a complementary tool, which enables to probe the excitations corresponding to the magnetic, lattice and electronic degrees of freedom. \cite{Lawes, Kadlec, Kamba, Vermette,Balkanski}. Temperature variation of Raman data substantiates the role of spin-phonon coupling associated with the Morin transition and its influence on the dielectric properties of $\alpha$- Fe$_2$O$_3$ samples investigated in this work.

\section{Experimental Techniques}

        \begin{figure*}[!t]
\includegraphics[width=1\textwidth]{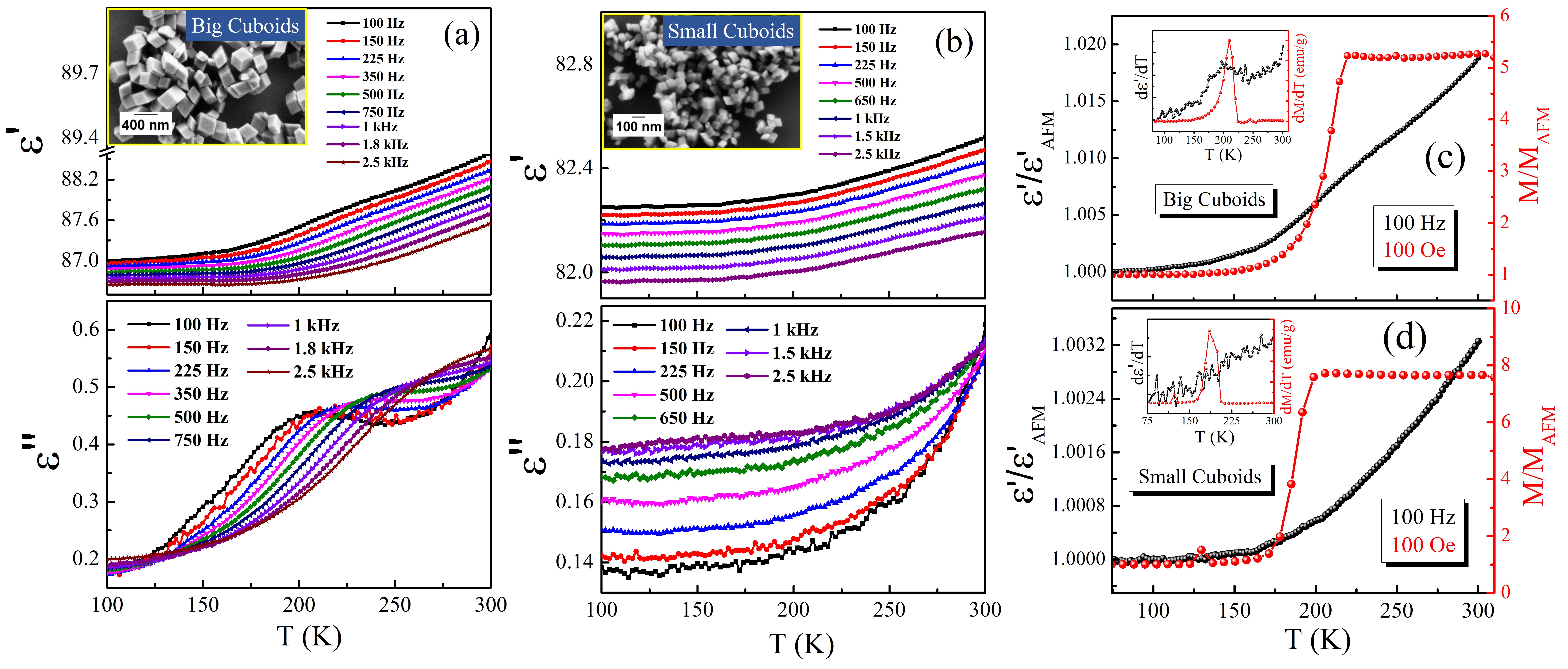}
\caption {Temperature variation of $\epsilon'$ (upper panel) and $\epsilon''$ (lower panel) in (a) Big Cuboids, (b) Small Cuboids. (c) an (d) are normalized  $\epsilon'$ in conjunction with the normalize Magnetization data for the big and small cuboids respectively. The insets in (c) and (d) depicts the temperature derivatives of $\epsilon'$ and magnetization. The inset in (c) reveals dielectric anomaly occurring in the vicinity of T$_M$ in big cuboids. }
\label{Figure2}
\end{figure*}

Big cuboids (side lengths $\sim$200 nm), small cuboids (side lengths $\sim$ 60 nm), micro plates (side length x thickness $\sim$ 1-3 µm x 300 nm) and nano plates (side length x thickness $\sim$ 70 nm x 15 nm) of $\alpha$- Fe$_2$O$_3$  have been synthesized by the hydrothermal method \cite{Pattanayak2}. The morphology of the  samples have been analysed by the Zeiss Ultra plus Field Emission Scanning Electron Microscope (FESEM) (insets of Figs. 1 and Figure 2). The SEM images show the cuboids are regular in shape with well-defined facets.  The plate shaped $\alpha$- Fe$_2$O$_3$  samples are nearly hexagonal in shape. The phase formation and crystallinity of the samples have been analysed by using a Bruker D8 advance powder X- ray diffractometer with Cu Kα radiation. The diffraction patterns fitted with Rietveld refinement using FULLPROF software are provided as SI1. Temperature dependent magnetization data recorded using a MPMS XL SQUID magnetometer while cooling the samples from room temperature down to 5 K in presence of 100 Oe field. The  bulk Morin transition is $\sim$ 250 K  $\alpha$- Fe$_2$O$_3$  and it is known to systematically decrease with the increase of S/V ratio, as we also observe in our samples\cite{Pattanayak2}. The surface to volume (S/V) ratio, Morin temperature and lattice constants extracted from the Rietveld refinement are presented in table-1. Temperature dependent Raman spectra has been acquired by using HORIBA LabRAM with the help of Linkam Stage.The Raman spectra is recorded using a green (Helium-Neon) laser of wave length 532 nm.  For the dielectric measurements,  as-prepared  powders of Hematite are  mixed  with a suitable binder, PVA (poly vinyl alcohol) and pelletized. These pellets, about 10 mm in diameter and 2 mm thickness are  sintered at a temperature of 400 C for 5 hours prior to dielectric measurements. The dielectric measurements  in parallel plate geometry  are carried out at a drive of 1 V from room temperature down to 75 K in the frequency range of 100 Hz - 300 kHz using a Novocontrol impedance analyzer. 
 
\section{ Results and Discussions}

 \begin{figure}[!t]
\includegraphics[width=0.5\textwidth]{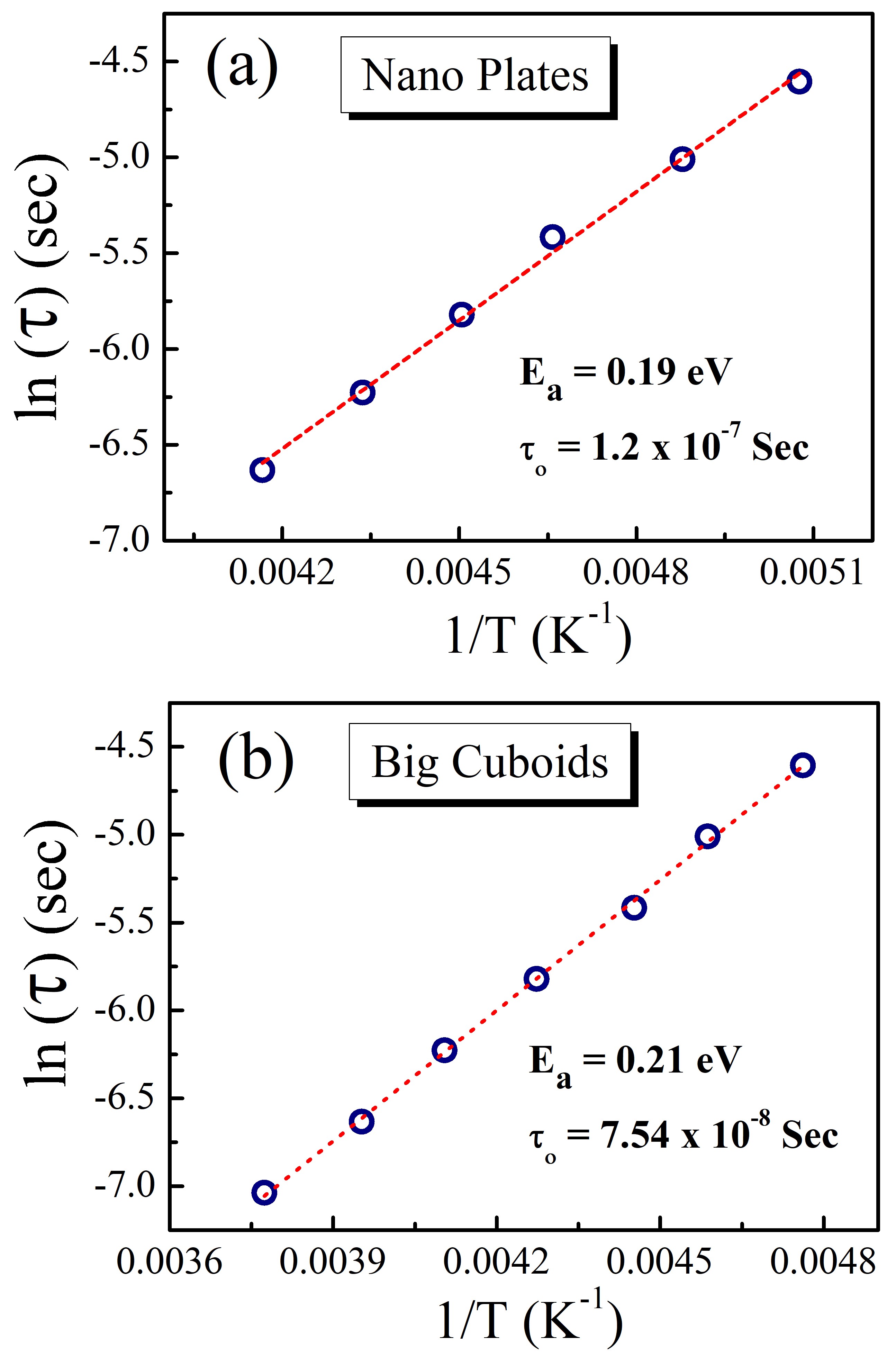}
\caption {Arrhenius plot of the relaxation time for the nano plates and big cuboids.}
\label{Figure3}
\end{figure}

\textcolor[rgb]{0,0,1}{\subsection{ Temperature variation Dielectric spectroscopy}}

The temperature dependent dielectric constant measured as a function of  frequency for micro- and nano plates is shown in Figs 1a-1b. SEM images in the upper figures reveal the morphology and size of the individual crystallites for both the samples under investigation.  The main panel of the figures 1a-1b represent the real ($\epsilon'$) and the imaginary part ($\epsilon''$) of the dielectric constant respectively. We note that magnitude of the  $\epsilon$ is nearly half ($\epsilon'$$\sim$ 60) in nano plates as compared to what is observed in micro-plates ($\epsilon'$$\sim$130). It is also to be noted that the Surface to Volume ratio ( SI-Table S1) for nano-plates is roughly 20 times larger than that of the micro-plates and the absolute value of dielectric constant is nearly half in this case. We also note that in both the samples, the dielectric constant exhibits a modest but clear increase in the WFM region. This feature is better seen when normalized $\epsilon'$ is plotted with normalized magnetization for both the samples (Figure 1c and Figure 1d). The anomaly in the vicinity of T$_M$ is more pronounced in micro plates as compared to nano plates, as is evident from comparing the derivatives (inset of Figure 1c). The magnitude of $\epsilon''$ is  low and it is of similar order, in the vicinity of T$_M$ in both cases. Interestingly, the $\epsilon''$ in case of  nano plates shows relaxation behavior with the maxima of dispersion shifting to higher temperature with the increase of frequency.  However,  micro plates do not show any  relaxation behavior in this frequency range. 

Figure 2 shows the same in case of small and big cuboids. Here the magnitude of $\epsilon'$ is rather  similar for both the samples (the small cuboids exhibit a marginally smaller value, $\epsilon'$$\sim$82 as compared to big cuboids, $\sim$88), even though  the S/V ratio for small cuboids is about 10 times that of big cuboids. The  over all trend is somewhat different from samples with hex plate morphology wherein $\epsilon'$ reduces significantly upon nano scaling. However big cuboids also exhibit a small anomaly in the vicinity of the Morin transition as is evident from the inset of Figure 2c wherein temperature derivatives of $\epsilon'$ and M are compared. Main Panel of Figures 2c and 2d (in which normalized $\epsilon'$ and normalized M is plotted) show increase in $\epsilon'$ in the WFM region for both the samples. We  also note that the $\epsilon''$ is similar in magnitude for big and small cuboids but the relaxation like feature is observed only in big cuboids (lower panel of Figure 2b). Overall, these data suggest  that both morphology and S/V ratio play a part as far as magnitude of  the $\epsilon'$ is concerned, but the  loss part, $\epsilon''$ is of similar range in all four cases , irrespective of morphology and S/V ratio. We also note that the S/V ratio of nano plates and big cuboids is of similar order and both exhibit relaxation type of behavior. The frequency range of this shift is in rather limited, up to a few kHz. This relaxation is seen to follow Arrhenius Law, given by the following equation.

\begin{equation}
\tau = \tau_0 e^{\frac{E_a}{k_BT}} 
\end{equation}

Here k$_B$ is the Boltzmann constant, T is the absolute temperature, $\tau_0$ is the pre exponential factor and E$_a$ is the activation energy of relaxation. The linear behavior observed following Arrhenius law is shown  for nano plates and big cuboids  in Figure 3a and 3b respectively. This is suggestive of thermally activated dipole population \cite{Thongbai,Zhang,Smart,Sharma1}. The activation energy deduced from the fitting for small cuboids  as well as nano-plates is $\sim$ 0.2 eV. Though, with present set of data,  it is non trivial to isolate its origin but this activation range is suggestive of polaron hopping, which is a plausible mechanism in case of hematite \cite{Smart}. Extrinsic contributions such as space charge or Maxwell-Wagner effect also does not seem to play a significant part, considering that magnitude of $\epsilon'$ and $\epsilon''$,  as these contributions tend to enhance both $\epsilon'$ and $\epsilon''$ \cite{Lunkenheimer, Liu, Catalan,Wang,Ke}. A small upturn in $\epsilon''$ curves observed in both the samples at higher temperature can be attributed to an extremely small change in electrical conductivity in that temperature region, which has also been reported for hematite. 

\begin{figure*}[!t]
\includegraphics[width=1\textwidth]{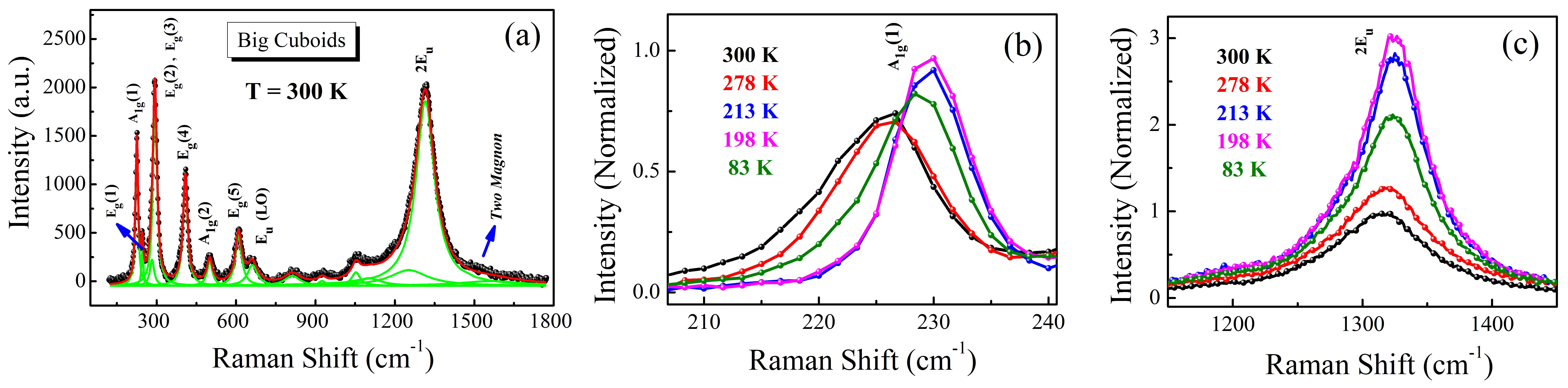}
\caption {Characteristic Raman spectra of  $\alpha$- Fe$_2$O$_3$ Big cuboids. Raman spectra as a function of temperature for (b) A$_1g$ (c) E$_g$ mode of vibrations. These data depict anomalous Raman shift in the modes on the either side of the Morin Transition.}
\label{Figure4}
\end{figure*}

Overall, from frequency dependent dielectric spectroscopy, we infer that  the modest rise in the dielectric constant is indeed tied to the onset of WFM region in all four samples, irrespective of the morphology and size.  This is also consistent with previous reports, exhibiting higher rise in the dielectric constant in the vicinity of FM to PM  transition as compared to what is observed in pure AFM to PM transition\cite{Lawes}. Thus, in case of hematite, it appears that the onset of DMI driven WFM has a role to play, and data is also  consistent with magnetodielectric coupling.  It is known that in the  MD systems, the dielectric anomaly appears at the magnetic transition does not necessarily evolve with the appearance of spontaneous electrical polarization. In such systems the appearance of dielectric anomaly at the vicinity of magnetic phase transition is governed by underlying  spin-spin correlations, which could be both FM or AFM \cite{Lawes}. This also enables to understand  subtle difference between AFM and FM correlations through the nature of spin spin correlations. As evident from Figure 3,  in case of $\alpha$- Fe$_2$O$_3$, irrespective of the morphology, the modest rise in the dielectric constant in each sample is linked to the onset of the WFM phase. It is also  known that in magnetic ferroelectrics, a strong polarization at the vicinity of the magnetic ordering resulted in the appearance of dielectric anomaly across the magnetic transition. In such systems a particular non collinear spin structure is an essential ingredient for the appearance of electrical polarization \cite{Aoyama,Kimura1,Hur,Katsura,Lawes}. Microscopically, the coupling between magnetic and electric order parameters in such systems is mainly governed by DMI interactions. Thus it is important to investigate the role of DMI, especially spin -phonon coupling in these samples. For this purpose we investigated temperature variation of Raman in a representative sample, big cuboids, which has also shown relaxation type of behavior in dielectric loss, which is discussed in next sub section.

\begin{figure*}[!t]
\includegraphics[width=1\textwidth]{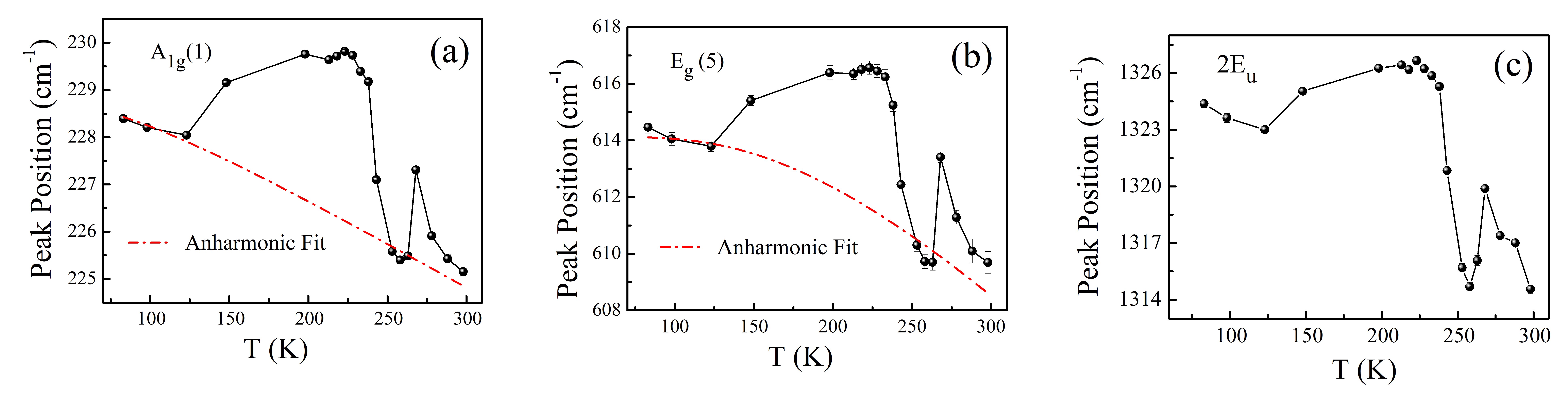}
\caption {Variation of peak positions with temperature for
the selected Raman modes (a) A$_1g (1)$, (b) E$_g (5)$, (c) E$_u$. The peak positions are extracted from the Lorentizian peak profile fit.}
\label{Figure5}
\end{figure*}

\textcolor[rgb]{0,0,1}{\subsection{ Temperature Variation of Raman }}

A characteristic Raman spectra acquired using green laser is shown in Fig.3 (a) on big cuboids of  $\alpha$- Fe$_2$O$_3$ sample  is shown in Fig. 4(a).  The assignment of Raman active modes is consistent with the group theory prediction for the space group R3-c \cite{Shim}. The modes located at 225 $cm^{-1}$ and 500 $cm^{-1}$ are the symmetric A$_1g$ modes and those located at 246 cm-1, 290 $cm^{-1}$, 295 $cm^{-1}$, 408 $cm^{-1}$, 500 $cm^{-1}$ and 610$cm^{-1}$are the doubly degenerate E$_g$ modes. The mode E$_u$, assigned at 660 cm-1 in general is not Raman active and its appearance has been attributed to the disorder related to surface defects and grain sizes. The mode located at 1320 $cm^{-1}$ is related by a factor of two with the E$_u$ mode. A weak peak observed in the spectra range of 1500-1600 $cm^{-1}$, merged with the tail of 2E$_u$ mode, is due to the two-magnon scattering in $\alpha$- Fe$_2$O$_3$  \cite{Rodriguez, Massey}. It is to be recalled that the A$_1g$ symmetry involves the movement of Fe atom along the c-axis of the unit cell and E$_g$ symmetry involves the symmetric breathing mode of the O atoms relative to the Fe atoms in the plane perpendicular to the c axis of the unit cell \cite{Beattie, Chernyshova}.

The temperature dependent Raman spectra acquired in the temperature range of 300 K - 80 K is shown in Fig. 3(b). In figures 3(c) and (d) the evolution of phonon frequency as a function of temperature, obtained after the Lorentzian peak profile fit of the acquired spectra, are shown for two selected modes A1g (1) and Eg (5) respectively. The variation of phonon frequency as a function of temperature in a magnetic material can be expressed as \cite{Massey}\\
      
     \begin{equation}
\begin{array}{r c l}
\omega(T) - \omega(0) = \Delta\omega(T) &=& \Delta\omega_{lattice}\\
 & & + \Delta\omega_{phonon-phonon} \\
& & + \Delta\omega_{spin-phonon}
\end{array}
\end{equation} 

Where $\omega$(T) and $\omega$(0) in L.H.S are the Raman frequency at T and 0 K respectively. The contribution from the first term in R.H.S is due to the lattice expansion/contraction, second term is due to phonon-phonon interaction and last term implies the contribution of spin-phonon coupling in the modulation of phonon frequency. Neglecting the contributions from the lattice and spin – phonon coupling, the variation of phonon frequency as a function of temperature can be simulated by the anharmonic decay of phonon frequency model expressed as \cite{Vermette, Balkanski}\\

\begin{equation}
\omega(T) = \omega(0) - C\left( 1 + \frac{2}{e^x - 1}\right) 
\end{equation}

Here C is an adjustable parameter, x = ћ $\omega$(0)/ k$_B$T, k$_B$ is the Boltzmann constant and T is the absolute temperature. The simulated curves of the anharmonic phonon decay are shown in red lines in Figures 4a and 4b. The phonon frequency evolutions as a function of temperature in three  modes follows the anharmonic behaviour (red lines) except a noticeable deviation observed around the Morin transition. Considering the subtle changes in lattice parameters around the Morin transition observed earlier in these samples, the deviations in the mode positions around the Morin transition can be attributed to arise due to the combined effect of spin-phonon and lattice contributions. We also observe a small deviation in phonon frequency evolution around the temperature regime of $\sim$ 250 K. We infer that this small deviation is more likely due to the presence of a double transition as evident from the  remanent magnetization vs temperature data reported earlier on this sample \cite{Pattanayak2}.The Raman data thus confirm the role of spin phonon coupling which manifests in the  dielectric measurements in hematite near AFM to WFM transition.
 
\section{Conclusions}
Temperature  and frequency dependent dielectric and Raman measurements  have been conducted around the Morin transition in  $\alpha$- Fe$_2$O$_3$ crystallites  formed in different morphologies and sizes. Nano scaling results in significant decrease  in the dielectric constant in case of hexagonal shaped plates whereas in case of cuboids, no significant changes were observed, as far as the magnitude of the dielectric constant is concerned. The real part of the dielectric constant exhibits a small but clear  increase in the weak ferromagnetic region as compared to pure antiferromagnetic region in all the samples, irrespective of morphology and size. This indicates presence of magnetodielectric coupling in hematite and also  bring forward the role of Dzyaloshinskii Moriya Interaction driven weak ferromagnetic phase. The temperature variation of Raman modes confirms the presence of spin phonon coupling at the vicinity of Morin transition.  

\section{Acknowledgments}
Authors thank Sunil Nair (IISER Pune) for SQUID magnetization measurements. AB acknowledges Department of Science and Technology (DST), India for funding support through a Ramanujan Grant and the DST Nanomission Thematic Unit Program.  

\bibliography{Bibliography}

\end{document}